\documentclass[full paper,twocolumn]{jpsj2}

\title{%
Quantum Spin Pump in $S=1/2$ antiferromagnetic chains \\
-Holonomy of phase operators in sine-Gordon theory-
}

\author{%
Ryuichi Shindou\thanks{E-mail address: shindou@appi.t.u-tokyo.ac.jp.}
}

\inst{%
Department of Applied Physics, University of Tokyo,
Bunkyo-ku, Tokyo 113-8656, Japan \\
RIKEN (The Institute of the Physical and Chemical
Research), 2-1 Hirosawa, Wako, Saitma 351-0198, Japan
}

\recdate{\today}

\abst{%
In this paper, we propose the quantum spin pumping in 
quantum spin systems where an applied electric 
field ($E$) and magnetic field ($H$) cause a finite spin gap to its 
critical ground state. When these systems are subject to 
alternating electromangetic fields; 
$(E,H)=(\sin\frac{2\pi t}{T},\cos\frac{2\pi t}{T})$ and travel 
along the {\it{loop}} $\Gamma_{\rm{loop}}$ which encloses their critical 
ground state in this $E$-$H$ phase diagram, the locking 
potential in the sine-Gordon model slides and changes its minimum.  
As a result, the phase operator acquires $2\pi$ holonomy during 
one cycle along $\Gamma_{\rm{loop}}$, which means 
that the quantized spin current has been transported 
through the bulk systems during this adiabatic process. 
The relevance to real systems such as 
Cu-benzoate and ${\rm{Yb}}_4{\rm{As}}_3$ is also discussed.  
}

\kword{%
field induced spin gap systems, quantized spin transport, 
Dirac monopole, sine Gordon theory, phase operator, 
topological stability, Landau Zener tunneling
}

\begin{document}
\maketitle
\section{Introduction}
Transport phenomena in magnets such as the 
colossal magnetoresistance\cite{tokura} 
and anomalous Hall effect\cite{kl,smit,nozieres, kondo,matl,chun,ye} 
are long-standing subjects in solid state physics. 
Recently, exotic characters of the anomalous 
Hall effect are closed up in ferromagnets\cite{ohgushi,taguchi,
kageyama,sonoda}, antiferromagnets\cite{shindou} 
and spin glass\cite{tatara,Kawamura} with non-coplanar spin 
configurations and collinear ferromagnets with strong spin-orbit 
couplings\cite{monoda,jungwirth,burkov,fang,yao}. 
Their common origin is gradually recognized as the topological 
character of magnetic Bloch wavefunctions in its ordered phase, 
sharing the same physics with the two dimensional ($2$D) 
quantum Hall effect. Thereby the spontaneous Hall conductivity 
is related to the {\it{fictitious magnetic field}}  defined 
in the crystal momentum space and its quantized value is directly 
related to the first Chern number associated with the fiber 
bundle whose base space is spanned by crystal momenta.
  
Meanwhile, in early 80's, Thouless and Niu proposed the 
quantized adiabatic particle transport (QAPT) in gapped fermi 
systems,  where an adiabatic sliding motion of the periodic 
electrostatic potential in one dimensional ($1$D) system pumps 
up an integer number of electrons per cycle.\cite{thouless,niu} 
In fact, this quantized particle transport is another physical 
manifestation of the topological character associated with  the Bloch 
wavefunctions. Specifically this particle (polarization) current 
is related to the fictitious magnetic field defined in 
the generalized crystal momentum space which is in turn spanned 
by the crystal momentum and {\it{deformed parameters}}. 
Because of this analogy to the 2D quantum Hall currents, this $1$D 
quantized particle current is known to be topologically 
protected against other perturbations such as disorders and 
electron-electron correlations\cite{niu}. Due to this peculiar 
topological protection, the QAPT became recently highlighted in 
the realm of the spintronics,  where electron pumpings 
have been experimentally realized in meso and/or nano-scale 
systems.\cite{altshuler,switkes,shilton,talyanskii} 
However the quantized value observed in these experiments  
should be attributed to the Coulomb blockade 
\cite{altshuler,shilton,talyanskii} and is different form 
the aforementioned Thouless quantization, which is assured 
only in the thermodynamic limit.\cite{altshuler}

In this paper, based on the original idea of the QAPT, we 
propose the quantized {\it{spin}} transport (quantum spin pump) in 
the macro-scale magnets. There we discuss systematically the topological 
stability of its quantization. Specifically, we clarify in 
this paper the relation between its quantization and 
{\it holonomy} of the phase operator of sine-Gordon theory 
in $1$D quantum field theory.  By using this relation, 
its topological stability can be quantitatively judged 
by seeing whether the expectation value of this phase operator 
acquires $2\pi$ holonomy during the cyclic process or not. 

As for the experimental relevant systems to our theory of quantized 
spin transports, we have the $S=\frac{1}{2}$ quantum 
spin chain such as Cu-benzoate and ${\rm{Yb}}_{4}{\rm{As}}_{3}$ 
(charge ordered phase), where its unit cell contains {\it{two}} 
crystallographically inequivalent sites. 
Accurately speaking, both the translational 
symmetry by one site  $(T:\mbox{\boldmath{$S$}}_j\rightarrow
\mbox{\boldmath{$S$}}_{j+1})$  
and the bond-centered 
inversion symmetry which exchanges the nearest
neighboring {\it{sites}} $(I_{\rm{bond}}: \mbox{\boldmath{$S$}}_{i-j}
\leftrightarrow \mbox{\boldmath{$S$}}_{i+j+1}$ 
are crystallographically broken. As a result of this peculiar 
crystal symmetry, the $g$-tensors of its two sublattices 
are in general different in these systems; 
\begin{eqnarray}
&&H=J\sum_{i=1}^{N}\mbox{\boldmath{$S$}}
_{i}\cdot\mbox{\boldmath{$S$}}_{i+1} + \nonumber \\ 
&&\hspace{0.8cm}\sum_{i}\mbox{\boldmath{$H$}}
\cdot\bigl[[g_{\rm{u}}] + (-1)^i[g_{\rm{a}}]
\bigr]\cdot\mbox{\boldmath{$S$}}
_{i},\hspace{0.2cm} (J>0) \label{-1-0} 
\end{eqnarray}
where $[g_{\rm{u}}]$ and $[g_{\rm{a}}]$ represent the uniform 
and staggered component of the $g$-tensors respectively.\cite{comm0} 
Late 90's, these quantum spin systems,
 whose ground states are    
nearly critical because of their strong one dimensionalities, are 
experimentally revealed to show the spin gap  behaviors under 
the external magnetic field. Furthermore, following 
theoretical works showed that the field-induced gap 
behaviors $(\Delta\sim H^{2/3})$ are originated from this 
staggered component of the $g$-tensor, where the effective staggered 
magnetic field in the eq. (\ref{-1-0}) endows its critical ground 
state with a finite spin gap.\cite{oshikawa,oshikawa1} 

On the other hand, the exchange interaction $J$ 
in these $S=1/2$ quantum spin chains, 
$J\sum_{i}\mbox{\boldmath{$S$}}_{i}\cdot\mbox{\boldmath{$S$}}_{i+1}$, 
does not  have an alternating component,  
since these systems are invariant 
under the site-centered inversion symmetry 
which exchanges the nearest neighbor {\it{bonds}} 
($I_{\rm{site}}:\mbox{\boldmath{$S$}}_{i-j}\leftrightarrow
\mbox{\boldmath{$S$}}_{i+j}$).
 However, when we break this symmetry by applying an electric 
field $\mbox{\boldmath{$E$}}$ along an appropriate direction\cite{comm2}, 
the exchange interaction in general does acquire 
a {\it staggered component};  
\begin{eqnarray}
H = \sum_{i=1}^{N}\bigl[J-(-1)^{i}\Delta\bigr]
\mbox{\boldmath{$S$}}_{i}\cdot
\mbox{\boldmath{$S$}}_{i+1}. \label{-1-0-1} 
\end{eqnarray}
Futhermore a site-centered inversion operation 
{\it{with the sign of $\mbox{\boldmath{$E$}}$ reversed}} requires that 
$\Delta$ must be an odd function of $\mbox{\boldmath{$E$}}$. 
This dimerizing field $\Delta$ is also expected to induce a finite 
mass to its critical ground state, causing the spin-Peierls state. 

Based on these observations, we propose a method of generating 
the quantized spin current in this type of spin systems 
($T, I_{\rm{bond}}$ : broken, $I_{\rm{site}}$ : unbroken) 
by using electromagnetic fields. Specifically, we study a 
following  $S=1/2$ Heisenberg model with time-dependent 
staggered field and bond alternation:
\begin{eqnarray} 
\hat{H}(t)&=& \sum_{i=1}^{N}\bigl[J-(-1)^{i}\Delta(t)\bigr]
\mbox{\boldmath{$S$}}_{i}\cdot\mbox{\boldmath{$S$}}_{i+1} \nonumber \\
&&\  +\ h_{\rm{st}}(t)\sum_{i=1}^{N}(-1)^{i}S^{z}_{i}. 
\label{-1-0-1-1} 
\end{eqnarray}
where staggered Zeeman field $h_{\rm st}(t)$ and bond alternations 
$\Delta(t)$ are supposed to be controlled by the applied 
electromagnetic fields. In $\S$ 2, we will argue by using
bosonization technique that the quantized number of $z$-component 
of spins are transported from one end of the system to the other 
during one cycle along the loop which encloses the critical ground 
state at $(\Delta,h_{\rm st})=(0,0)$ in the 
$\Delta$-$h_{\rm st}$ plane.

In order to uphold this bosonization argument, we also 
demonstrate the numerical calculations 
in $\S$ 3, where we evolve the ground state 
wavefunction according as the above time-dependent 
Hamiltonian. Thereby we attibute the quantized spin 
transport to the Landau-Zener tunnelings \cite{zener} 
 which indeed happen between several energy levels during this 
cycle process. 

In addition to eq.(\ref{-1-0-1-1}), we also have a uniform 
component of Zeeman fields in real systems,
\begin{eqnarray}
H'(t) = \sum_{j=1}^{N}\mbox{\boldmath{$H$}}(t)\cdot 
\mbox{\boldmath{$S$}}_{j}, \nonumber 
\end{eqnarray}  
whose magnitude is usually larger than that of the staggered 
component $h_{\rm st}(t)$. This uniform 
field, especially its $x$ or $y$ component, 
seems to easily flip the spins accumulated at both 
boundaries and might spoil the physical consequence of the spin 
transport. Then we also discuss in $\S$ 3 the effect of 
this uniform field and the remedy  against it, where it turns 
out that the sweeping velocity should be appropriately 
chosen so as to avoid the relaxation process via this  
uniform Zeeman field.

\section{Bosonization}
For clarity, let us first illustrate the physics of our 
quantum spin pumping in a simple limiting case; 
\begin{eqnarray} 
H_{\rm{XY}}&=& -\frac{J}{2}\sum_{i=1}^{N}
(\hat{S}^{+}_{i}\hat{S}^{-}_{i+1} + {\rm{c.c.}}),\label{1-0-1} \\  
H_{\rm{dim}}&=& \frac{\Delta}{2}\sum_{i=1}^{N}(-1)^i
\left(\hat{S}^{+}_{i}\hat{S}^{-}_{i+1}
+\hat{S}^{-}_{i}\hat{S}^{+}_{i+1}\right), \label{1-0-2} \\
H_{\rm{st}}&=&h_{\rm{st}}\sum_{i}(-1)^{i}\hat{S}^{z}_{i},\label{1-0-3} 
\end{eqnarray}
where we take the direction of the staggered Zeeman field as the 
$z$-direction.\cite{comm1-1} In this simplification, 
we neglected 
the exchange coupling between the $z$-component of 
spins.  In this limit, we can get a quadratic form of 
Hamiltonian in terms of the Jordan-Wigner (JW) fermion 
$\hat{S}^{z}_{i} = f^{\dagger}_{i}f_{i}-\frac{1}{2},
\hat{S}^{+}_{i} = f^{\dagger}_{i}\exp({{\rm i}\pi\sum_{j=1}^{i-1}
f^{\dagger}_{j}f_{j}})$\cite{jw,lsm}. Then $H_{\rm{XY}}$ 
forms a cosine band 
in its momentum space, $\epsilon(k)=-J\cos k\alpha$ ($k$,$\alpha$ are
crystal momentum and lattice constant). In the absence 
of $\Delta$ and $h_{\rm{st}}$, the fermi points locate 
at $k=\pm\frac{\pi}{2\alpha}$ since fermions in the ground state 
fill up all the $k$ points but those with positive 
energy $\epsilon(k)$.

Nonzero $\Delta$ and/or $h_{\rm{st}}$ introduce a finite gap 
at these two Fermi points. The dimer state induced by a finite $\Delta$ 
can be understood as a {\it{Peierls insulator}}, where the JW fermions 
occupy the bonding orbitals between the two 
neighboring sites and form a valence band in 
its $k$-space. On the contrary, the antiferromagnetic 
state induced by the effective staggered magnetic 
field $h_{\rm{st}}$ along the $z$-direction 
can be interpreted as an {\it{ionic insulator}}, where 
the JW fermions stay on every other site. 
The conduction band and the valence band 
touch at $k=\pm\frac{\pi}{2\alpha}$, 
only when $(\Delta,h_{\rm{st}})$ is taken at the origin in this 
$\Delta\mbox{-}h_{\rm{st}}$ plane. Because of the periodicity of 
$\frac{\pi}{\alpha}$ along the $k$-axis, the double degeneracy point 
at $(k,\Delta,h_{\rm{st}})=(\frac{\pi}{2\alpha},0,0)$ is identical 
to that of $(k,\Delta,h_{\rm{st}})=(-\frac{\pi}{2\alpha},0,0)$.
 
Elementary analyses show that this double degeneracy point 
becomes the source (sink) of the vector field 
$\mbox{\boldmath{$\cal B$}}_{+1}$ ($\mbox{\boldmath{$\cal B$}}_{-1}$)  defined 
in the generalized crystal momentum space 
($k$-$\Delta$-$h_{\rm{st}}$ space); 
\begin{eqnarray}
\mbox{\boldmath{$\cal B$}}_{n}(\mbox{\boldmath{$K$}})
&=&\mbox{\boldmath{$\nabla$}}_{\mbox{\boldmath{$K$}}}
\times\mbox{\boldmath{$\cal A$}}_{n}(\mbox{\boldmath{$K$}}),
\nonumber \\
\mbox{\boldmath{$\cal A$}}_{n}(\mbox{\boldmath{$K$}}) 
&=&\frac{\rm i}{2\pi}\langle n (\mbox{\boldmath{$K$}}) |
 \mbox{\boldmath{$\nabla$}}_{\mbox{\boldmath{$K$}}} |
 n (\mbox{\boldmath{$K$}})\rangle, \nonumber
\end{eqnarray}  
where $\mbox{\boldmath{$K$}}=(k,\Delta,h_{\rm{st}})$ 
and $\mbox{\boldmath{$\nabla$}}_{\mbox{\boldmath{$K$}}}
=(\partial_{k},\partial_{\Delta}
,\partial_{h_{\rm{st}}})$. Here 
$|n(\mbox{\boldmath{$K$}})\rangle$ represents the periodic part of 
the Bloch function for the valence band ($n=+1$) and 
the conduction band ($n=-1$).  The vector field ${\cal A}_{n,\mu}$  
changes by $\frac{1}{2\pi}\nabla_{K_{\mu}}\phi$ under the U(1) gauge 
transformation of these Bloch wavefunctions; 
$|n(\mbox{\boldmath{$K$}})\rangle \rightarrow \exp[{\rm i}\phi(
\mbox{\boldmath{$K$}})]|n(\mbox{\boldmath{$K$}})\rangle$, 
while its rotation, i.e., $\mbox{\boldmath{$\cal B$}}_{n}$ remains invariant.
Because of this gauge invariance,  
we often call the latter vector field $\mbox{\boldmath{$\cal B$}}_{n}$ 
as a {\it{fictitious magnetic field}} 
or flux. Correspondingly, %
the double degeneracy point located at 
$(\pm\frac{\pi}{2\alpha},0,0)$ will be referred to as 
the {\it{fictitious magnetic charge}} 
(magnetic monopole in the generalized momentum space), 
whose magnetic unit can be shown to be $1$:  
\begin{eqnarray}
\int_{S_{1}} {\rm d}\mbox{\boldmath{$S$}}\cdot
\mbox{\boldmath{$\cal B$}}_{\pm 1}=\pm 1.\label{0-0} 
\end{eqnarray}
Here $S_{1}$ represents the arbitrary closed surface which 
encloses the double degeneracy point at $(k,\Delta,h_{\rm{st}})
=(+\frac{\pi}{2\alpha},0,0)\equiv(-\frac{\pi}{2\alpha},0,0)$.  

Based on these observations, let us consider the adiabatic 
process where the two parameters $(\Delta,h_{\rm{st}})$ are changed 
along a loop $\Gamma_{\rm{loop}}$ enclosing the origin
($(\Delta,h_{\rm{st}})=(0,0)$); 
\begin{eqnarray}
&&(h_{\rm{st}},\Delta) \equiv R(\cos\varphi,\sin\varphi) \label{0-0-1} \\
&&R\ne 0\ ,\ \ \varphi : 0 \rightarrow 2\pi. \label{0-0-2} 
\end{eqnarray}
Then, according to the original idea of the
QAPT,\cite{thouless,niu,avron}  the total number of 
 JW fermions ($I$) which are transported from one side 
of this system to the other (in the positive direction) 
during this adiabatic process is  
equal to the total flux for the valence 
band ($n=1$), $\mbox{\boldmath{$\cal B$}}_{+1}$, which penetrates 
the 2D closed sphere spanned by $\Gamma_{\rm{loop}}$ 
and the Brillouin zone :   
\begin{eqnarray}
I=\int_{\Gamma_{\rm{loop}}
\times[-\frac{\pi}{2\alpha},\frac{\pi}{2\alpha}]} 
{\rm d}\mbox{\boldmath{$S$}}\cdot\mbox{\boldmath{$\cal B$}}_{+1}=+1.
\end{eqnarray}
 
The physical meaning of this quantized  fermion 
transport is nothing but the quantized spin transport,  
since the JW fermion density is related to 
the $z$-component of the spin density.  
In other words, the total $S^{z}$ around one end of this system 
decreases by 1 while that of the other end increases by 1 during 
this adiabatic cycle. This quantization is topologically protected 
against the other perturbations as long as the gap along the 
loop remains finite\cite{niu,avron}, in other words, as far as 
the double degeneracy points do not get out of (enter into) the 2D 
closed surface $\Gamma_{\rm{loop}}\times[-\frac{\pi}{2\alpha},
\frac{\pi}{2\alpha}]$.  Then, we naturally expect that this 
quantized spin transport is stable against the weak exchange 
interactions between the $z$-components of spins;
\begin{eqnarray}
H_{\rm{Z}}&=&|\gamma|\sum_{j}\bigl[J+(-1)^{j}\Delta\bigr]\hat{S}^{z}_{j}
\hat{S}^{z}_{j+1}.\label{1-0-4} 
\end{eqnarray}

In the following, we will prove that this is indeed the case,  
by introducing the exchange interactions between $S^{z}$.
As a first step, we will treat $H_{\rm{XY}}$ term 
as a non-perturbed term and review which perturbations 
are relevant among  $H_{\rm{Z}}$, $H_{\rm{dim}}$ and $H_{\rm{st}}$ 
by using the bosonization technique. Namely, we 
first introduce the slowly varying fields, $R(x),L(x)$:  
\begin{eqnarray} 
f_{j}&\simeq 
& R(x_j)e^{{\rm i}k_{F}x_j}+L(x_j)e^{-{\rm i}k_{F}x_j}, \nonumber \\
R(x_j) &=& \sum_{|k-k_{F}|\ll \alpha^{-1}}
f_{k}e^{{\rm i}(k-k_{F})x_j},\nonumber \\
L(x_j) &=& \sum_{|k+k_{F}|\ll  \alpha^{-1}}
f_{k}e^{{\rm i}(k+k_{F})x_j}.\nonumber 
\end{eqnarray}
Then we rewrite the spin operator in terms of these fermion fields 
$R(x_j)$ and $L(x_j)$:  
\begin{eqnarray}
&&\hat{S}^{z}_{j}=f^{\dagger}_{j}f_{j}-\frac{1}{2}  \nonumber \\
&&\simeq:R^{\dagger}(x_{j})R(x_{j}):+:L^{\dagger}(x_{j})L(x_{j}): 
\nonumber \\
&&\ \  + (-1)^{j}\Bigl[R^{\dagger}(x_{j})L(x_{j})+L^{\dagger}(x_{j})
R(x_{j})\Bigr],\label{1-1-1} \\
&&\hat{S}^{+}_{j}\hat{S}^{-}_{j+1}
+\hat{S}^{-}_{j}\hat{S}^{+}_{j+1} 
=f^{\dagger}_{j}f_{j+1}+f^{\dagger}_{j+1}f_{j}\nonumber \\
&&\simeq {\rm i}\alpha\cdot\left[:R^{\dagger}(x_{j})
{\partial_x}R(x_{j}): - :L^{\dagger}(x_{j}){\partial_x}L(x_{j}): 
- {\rm{H.c.}}\right] \nonumber \\
&&\ \ -2{\rm i}(-1)^{j}\biggl[R^{\dagger}(x_j)L(x_j)
-L^{\dagger}(x_j)R(x_j)\biggr],\label{1-1-2} 
\end{eqnarray}
where $:...:$ stands for the normal order.\cite{NN} Here we retain 
the lowest order terms with respect to the lattice constant $\alpha$ 
both for the uniform and staggered components. 
According to the bosonization recipe,\cite{shankar} 
we can rewrite these fermion 
operators by using the phase operator $\hat{\theta}_{+}(x)$ and 
its canonical conjugate field $\hat{\Pi}(x)$. Namely, 
the spin operators in eqs.(\ref{1-1-1}) and (\ref{1-1-2})  
read;
\begin{eqnarray}
&&\hat{S}^{z}_{j}=\frac{\partial_{x}\hat{\theta}_{+}(x_{j})}{2\pi} - (-1)^{j}
\frac{1}{\pi\alpha}\sin\hat{\theta}_{+}(x_{j}), \label{2-0-1} \\
&&\hat{S}^{+}_{j}\hat{S}^{-}_{j+1}+\hat{S}^{-}_{j}\hat{S}^{+}_{j+1} 
\nonumber \\
&&\hspace{0.5cm}=-\alpha\Bigr[
\frac{1}{4\pi}{\bigl(\partial_{x} \hat{\theta}_{+}(x_{j})\bigr)}^2
 + 4\pi\hat{\Pi}(x_{j})^{2}\Bigl] \nonumber \\
&&\hspace{0.8cm} - (-1)^{j}\frac{2}{\pi\alpha}
\cos\hat{\theta}_{+}(x_{j}).\label{2-0-2} 
\end{eqnarray}
Then we substitute these equations into 
eqs.(\ref{1-0-1})$-$(\ref{1-0-3}) and (\ref{1-0-4}) and 
obtain the following expressions for 
$H_{\rm{XY}},H_{\rm{Z}},H_{\rm{dim}}$ 
and $H_{\rm{st}}$ in the continuum limit, 
where we neglect the rapid varying components such as 
$\sum_{j}(-1)^{j}\cos\hat{\theta}_{+}(x_j)$ and etc.: 
\begin{eqnarray}
&&H_{\rm{XY}}=\int {\rm d}x
\Biggl[\ 2\pi J\hat{\Pi}(x)^{2} + \frac{J}{8\pi}
(\partial_{x}\hat{\theta}_{+}(x))^{2} \Biggr],\label{2-0-3} \\
&&\hspace{0.2cm}H_{\rm{Z}}=\frac{|\gamma|}
{\alpha}\int {\rm d}x \Biggl[\frac{J}{4\pi^{2}} 
(\partial_{x}\hat{\theta}_{+}(x))^{2} + \frac{J}{2(\pi\alpha)^{2}}
\cos2\hat{\theta}_{+}(x) \nonumber \\
&& \hspace{1.1cm} -\frac{\Delta}{(\pi\alpha)^{2}}
\cos\hat{\theta}_{+}(x)\Biggr],\label{2-0-3-1} \\
&&H_{\rm{dim}}=-\frac{\Delta}{\pi\alpha^{2}}\int 
{\rm d}x\cos\hat{\theta}_{+}(x) \label{2-0-3-2} \\ 
&&\hspace{0.2cm}H_{\rm{st}} =-\frac{h_{\rm{st}}}{\pi\alpha^{2}}\int 
{\rm d}x\sin\hat{\theta}_{+}(x),\label{2-0-3-3} 
\end{eqnarray}
Here the origin of the third term of eq.(\ref{2-0-3-1}) were 
discussed elsewhere\cite{eggert}, which 
we will omit by redefining $\Delta$ in eq. (\ref{2-0-3-2}) 
in the followings. Then our final expression for Hamiltonian reads, 
\begin{eqnarray}
H&=&\int {\rm d}x \biggl\{v\Bigl[\pi\eta\hat{\Pi}^{2}
+\frac{1}{4\pi\eta}(\partial_{x}\hat{\theta}_{+})^{2}\Bigr] \nonumber \\
&-&\frac{R}{\pi\alpha^{2}}
\sin(\hat{\theta}_{+}+\varphi) 
+\frac{|\gamma| J}{2\pi^{2}\alpha^{3}}\cos2\hat{\theta}_{+} \biggr\},
\label{2-0-3-4}
\end{eqnarray}
where the velocity $v$ is given by  
$v=\frac{J}{2}\sqrt{(1+\frac{2|\gamma|}{\pi\alpha})}$.
We also introduced the quantum parameter  $\eta$ as 
\begin{eqnarray}
\eta=2\sqrt{\frac{1}{\bigl(1+\frac{2|\gamma|}{\pi\alpha}\bigr)}} 
\label{2-0-3-5}
\end{eqnarray}
which measures the strength of the quantum fluctuation. 
Namely, when $|\gamma|$ in eq.(\ref{1-0-4})
grows from $0$, this parameter decreases 
monotonically from 2. Even though 
eqs. (\ref{2-0-3})$-$(\ref{2-0-3-5}) were derived in the weak 
coupling limit ($|\gamma|\ll 1$), the final form of 
eq. (\ref{2-0-3-4}) is known to be valid until $|\gamma|$ 
reaches 1 (isotropic Heisenberg model), 
where the SU(2) rotational symmetry of correlation functions 
($\langle S^{x}_{j}S^{x}_{i}\rangle \sim 
\langle S^{z}_{j}S^{z}_{i}\rangle$) requires that 
$\eta=1$ at $|\gamma|=1$.\cite{NN} 
By using this quantum parameter $\eta$, 
the renormalization group (RG) eigenvalue of 
$\sin(\hat{\theta}_{+}+\varphi)$ is represented by $2-\frac{\eta}{2}$ 
while that of $\cos2\hat{\theta}_{+}$ is given by $2-2\eta$. 
This indicates that, as long as $|\gamma|\leq 1$ ($1\leq\eta\leq 2$) , 
the exchange coupling between the $z$-component 
of spins ($\frac{|\gamma| J}{2\pi^{2}\alpha^{3}}\cos2\hat{\theta}_{+}$) 
is always irrelevant in the sense of renormalization group 
analyses, while the dimerizing field $\Delta$ 
and staggered field $h_{\rm{st}}$ 
($\frac{R}{\pi\alpha^{2}} \sin(\hat{\theta}_{+}+\varphi)$)
 are equally relevant and 
lock the phase operator $\hat{\theta}_{+}$ on 
$\frac{\pi}{2}-\varphi+2\pi n$.  

\begin{figure}[t]
\begin{center}
\includegraphics[width=0.4\textwidth]{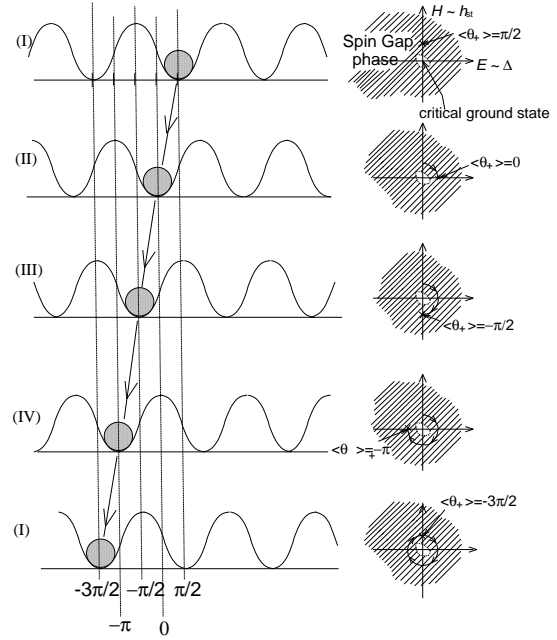}
\end{center}
\caption{A sine curve represents the locking potential which 
slides adiabatically. As long as the sliding speed is low enough, 
the system (shaded circle) stays the same valley and does not jump 
into its neighboring valleys.}
\label{1}
\end{figure}

As the system is always locked by  
$\sin(\hat{\theta}_{+}+\varphi)$ for $|\gamma|\leq 1$, 
we naturally expect that 
the quantized spin transport in the case of $|\gamma|=0$ could be 
generalized into the case of finite $|\gamma|$, at least up to $|\gamma|=1$.
This expectation is easily verified when we notice the physical 
meaning of the phase operator $\frac{\hat{\theta}_{+}}{2\pi}$.  
Since the spatial derivative of the phase operator corresponds 
to the $z$-component of spin {\it{density}}, this phase operator is   
 nothing but minus of the {\it{spatial polarization}} of the $z$-component 
of spins, i.e., $-\hat{P}_{S^{z}}\equiv-\frac{1}{N}\sum_{j=1}^{N}
j\hat{S}^{z}_{j}$. The equivalence between these two quantities is 
discussed in detail  in the appendix. 
Then, through the adiabatic process eqs.(\ref{0-0-1}) and(\ref{0-0-2}),  
$\langle \hat{\theta}_{+}\rangle$ decreases monotonically and 
acquires $-2\pi$ holonomy after one cycle (see
Fig. \ref{1}). In other words, $P_{S^{z}}$ 
increases by 1 per one cycle, i.e., 
\begin{eqnarray}
\delta P_{S^{z}}\equiv\oint_{\Gamma_{\rm{loop}}}{\rm d}P_{S^{z}}=-\frac{1}{2\pi}
\oint_{\Gamma_{\rm{loop}}}{\rm d}\mbox{\boldmath{$\lambda$}}\cdot
\partial_{\mbox{\boldmath{$\lambda$}}}
\langle\hat{\theta}_{+}\rangle=1.\label{4-1-0}
\end{eqnarray}
This relation always holds as far as the system is locked 
by the sliding potential $\sin(\hat{\theta}_{+}+\varphi(t))$, which 
is true for $|\gamma|\leq 1$.

Then we will argue the physical consequence of eq. (\ref{4-1-0}). 
Generally speaking, when the bulk system has a finite spin gap, 
the effects of boundaries range over its magnetic correlation length 
from the both ends. Then we naturally divide the contribution 
to $P_{S^{z}}$ into following two parts; 
\begin{eqnarray}
P_{S^{z}} &=& P^{\rm{in}}_{S^{z}} + P^{\rm{edge}}_{S^{z}},
\hspace{0.3cm} P^{\rm in(edge)}_{S^{z}} = \frac{1}{N}
\sum_{j\in{\Omega_{\rm in(edge)}}}
j\cdot \langle \hat{S}^{z}_{j}\rangle \nonumber  
\end{eqnarray}
where the $j$-summation in $P^{\rm{in}}_{S^{z}}$ are restricted within 
the interior of the systems ($\Omega_{\rm{in}}$), while spins 
within the edge region ($\Omega_{\rm{edge}}$) contribute 
to $P^{\rm{edge}}_{S^{z}}$. The ``interior'' of the system 
is defined as the  region where the spin state is same as that of 
the periodic boundary condition (p.b.c.). Namely, the presence of 
the boundaries does not influence on the spin 
state of the interior. On the other hand, the spin state within 
the ``edge'' region is strongly affected by the boundaries. 
Then, by definition, the spins within the interior 
would turn back to the same spin configuration as 
that of the initial state after an {\it{adiabatic}} 
evolution along $\Gamma_{\rm{loop}}$, 
\begin{eqnarray}
\delta P^{\rm{in}}_{S^{z}}=0.\nonumber 
\end{eqnarray} 
Therefore the difference of the spatial 
polarization of $\langle \hat{S}^{z}_{j}\rangle$ 
should be attributed to the change of spin 
configurations in the {\it{edge}} region,
\begin{eqnarray} 
\delta P^{\rm{edge}}_{S^{z}} = \frac{1}{N}
\sum_{j\in{\Omega_{\rm{edge}}}}
j\cdot \delta\langle \hat{S}^{z}_{j}\rangle = 1. \label{4-2+1}
\end{eqnarray} 

In eq.(\ref{4-2+1}), we next approximate $j$ for sites 
around the right end by $N/2$ and those of left by $-N/2$; 
\begin{eqnarray}
-\frac{1}{2}\sum_{j\simeq -\frac{N}{2}} \delta \langle S^{z}_{j} \rangle 
+ \frac{1}{2}\sum_{j\simeq \frac{N}{2}} \delta \langle S^{z}_{j} \rangle 
\approx 1, \label{4-2}
\end{eqnarray}
where we take the origin of site index $j$ at the center of the system. 
This approximation is allowed, since the edge region ranges over the 
magnetic correlation length from both ends, which is at most several 
sites in gapped spin systems. Thereby the semi-equal sign in
eq.(\ref{4-2}) could be safely replaced by the equal sign in the 
thermodynamic limit. When we bear in mind that 
the system has been staying on the eigenspace of  
$\hat{S}^{z}_{\rm{tot}}=0$ during this process; 
\begin{eqnarray}
\sum_{j\simeq -\frac{N}{2}} 
\delta \langle S^{z}_{j} \rangle  + \sum_{j\simeq \frac{N}{2}} 
\delta \langle S^{z}_{j} \rangle = 0, \nonumber 
\end{eqnarray} 
eq.(\ref{4-2}) indicates that the total $S^{z}$ 
around the right end ($j\simeq N/2$) increases 
by $1$ while that of the left end ($j\simeq -N/2$) decreases 
by $1$ along this cyclic evolution:
\begin{eqnarray}
\sum_{j\in {\rm left\ end}} 
\delta \langle S^{z}_{j} \rangle = -1, \hspace{0.3cm}
\sum_{j\in {\rm right\ end}} 
\delta \langle S^{z}_{j} \rangle = 1. \nonumber 
\end{eqnarray}
Namely, the spin current quantized to $1$ flows from the left end 
of the system to the right end during this one cycle. 

In summary, the quantized spin transport is always assured 
as long as $|\gamma|\le 1$.
This story does not alter even with the finite  
sliding speed $c=\frac{2\pi}{T}$,  as far as it is 
slower than the height of this locking potential, 
in other words, the spin gap along the loop.  
\section{Numerical Analyses}
In order to confirm the above argument, we performed the numerical 
calculation in the case of $|\gamma|=1$. Namely,  
we generated numerically the temporal evolution of the ground state 
wavefunction under the following time-dependent Hamiltonian: 
\begin{eqnarray}
&&|\phi(t)\rangle={\cal T}
\Bigl\{\exp[{\rm i}\int_{0}^{t}{\rm d}t^{\prime}\hat{H}(t^{\prime})]\Bigr\}
\cdot|g\rangle_{\rm{I}}, \nonumber \\
&&\hat{H}(t)= \sum_{i=1}^{N}\bigl[J-(-1)^{i}\Delta(t)\bigr]
\mbox{\boldmath{$S$}}_{i}\cdot
\mbox{\boldmath{$S$}}_{i+1} \nonumber \\
&&\hspace{1.0cm} +\ h_{\rm{st}}(t)\sum_{i=1}^{N}(-1)^{i}S^{z}_{i},
\nonumber \\
&&(h_{\rm{st}}(t),\Delta(t))=R(\cos \frac{2\pi t}{T}
,\sin \frac{2\pi t}{T}), \label{5-0} 
\end{eqnarray}
where $\cal{T}$ represents the time order operator, 
 $T$ is the cycle period during which the system has 
swept the loop ($\Gamma_{\rm{loop}}$) one time and $|g\rangle_{\rm{I}}$ 
is the ground state wavefunction at $(h_{\rm st},\Delta) = (R,0)$.

\begin{figure}[t]
\begin{center}
\includegraphics[width=0.4\textwidth]{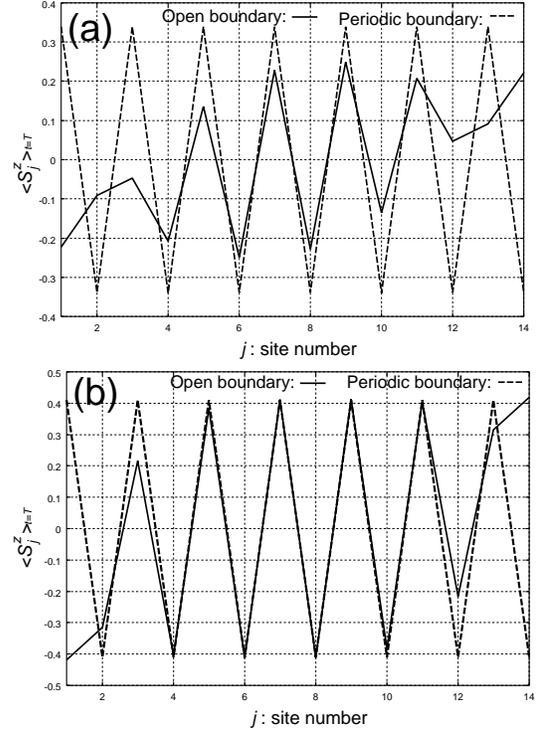}
\end{center}
\caption{(a): $\langle \hat{S}^{z}_{j} \rangle_{t=T}$ for  
 $J=1.5,R=0.3$ and $T=40$, where   
$\delta P_{S^{z}}\equiv\langle \hat{P}_{S^{z}} \rangle_{t=T}-
\langle \hat{P}_{S^{z}} \rangle_{t=0}=0.623$,  
(b): $\langle \hat{S}^{z}_{j} \rangle_{t=T}$ for 
$J=0.6,R=0.3$ and $T=80$, where   
$\delta P_{S^{z}}=0.862$.}
\label{2}
\end{figure}

In Fig. \ref{2}, we show the expectation value of $\hat{S}^{z}_{j}$ taken
over the final state, $|\phi(t=T)\rangle$, both with the 
p.b.c. and with the open boundary condition. 
The parameter $R$ is 
fixed  to be $0.3$ and $J$  is taken as $1.5\ (0.6)$ in Fig. \ref{2}a(b). 
The cycle period $T$ is taken as 40 
for the former and 80 for the latter, both of which are 
sufficiently long compared with the inverse of the spin gap
observed along the loop. 
 
For the p.b.c., as far as the sweeping velocity 
is low enough, the final state gives completely same 
configurations of  $\langle \hat{S}^{z}_{j}\rangle$  as that of the 
initial antiferromagnetic ground state. 
However in the case of the o.b.c., the spins around  
both boundaries are clearly modified as seen in Fig. \ref{2}. 
When we read the total $S^{z}$ around the left end ( 
$\sum_{j=1}^{3}\langle \hat{S}^{z}_{j}\rangle$ )
and that of the right end  ( 
$\sum_{j=12}^{14}\langle \hat{S}^{z}_{j}\rangle$ ) 
from Fig. \ref{2}a(b), the former 
increases by 0.67 (0.9) while the latter decreases by 0.67 (0.9)  
after this cyclic process. This result is qualitatively 
consistent with our preceding arguments. 

\begin{figure}
\begin{center}
\includegraphics[width=0.45\textwidth]{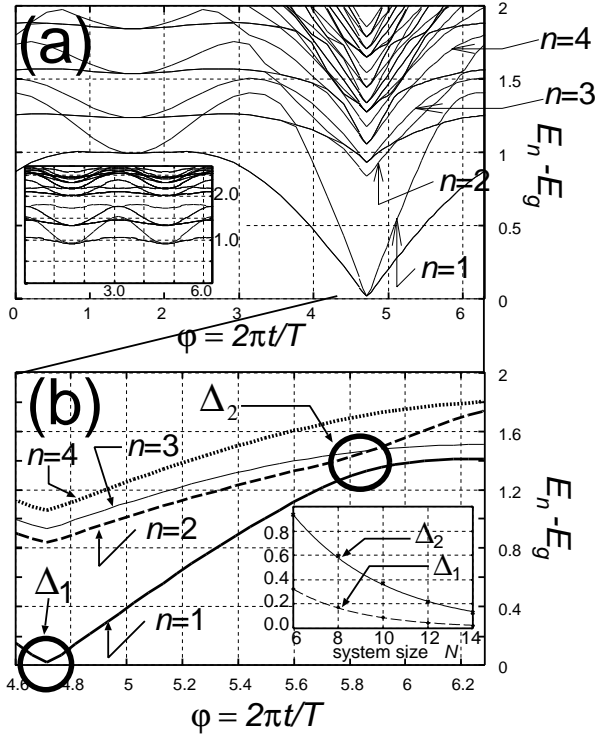}
\end{center}
\caption{(a): Energy levels of all the eigenstates 
as a function of $\varphi \equiv \frac{2\pi t}{T}$ 
for the o.b.c. (inset: p.b.c.). $J=1.5$, $R=0.3$ and $N=14$. 
(b): Energy levels of those eigenstates with $\hat{S}^{z}_{\rm{tot}}=0$
 for the o.b.c. $J=1.5$, $R=0.3$ and $N=14$.
(inset: Size-dependence of the minimum energy gaps 
$\Delta_{1}\equiv {\rm{min}}_{\varphi}\bigl[E_{n=1}(\varphi)-E_{g}(\varphi)
\bigr]$ and $\Delta_{2}\equiv {\rm{min}}_{\varphi}
\bigl[E_{n=2}(\varphi)-E_{n=1}(\varphi)\bigr]$. 
Fitting curves  for $\Delta_1$ and $\Delta_2$ 
are $0.32\exp(-(N-6)/3)$ and 
$0.925\exp(-(N-6)/4.3)$ respectively.)}
\label{3} 
\end{figure}

In order to understand the difference between 
the result of the o.b.c. and that of the p.b.c., we also calculated  
the instantaneous eigenenergy as a function of $\varphi\equiv \frac{2\pi
t}{T}$ in Fig. \ref{3}, where we take the offset as the ground state 
energy.
For the p.b.c. (inset of Fig. \ref{3}a)
, there is always a finite energy gap from 
the ground state for all $\varphi$. 
On the contrary, the gap for the o.b.c. reduces 
strongly at $\varphi=\frac{3\pi}{2}$ (Figs. \ref{3}a and \ref{3}b). 
Furthermore, this reduced energy gap $\Delta_{1}$ 
becomes smaller and smaller when we take the system size $N$ larger 
(inset of Fig. \ref{3}b, N=6,8,10,12), and the four states\cite{comm4} 
including the ground state are nearly 
degenerate at $\varphi=\frac{3\pi}{2}$ for $N=14$. 

Let us discuss the physical meanings of this quasi-degeneracy.
At  $(h_{\rm{st}},\Delta)=(0,-R)$, the exchange 
coupling between $\mbox{\boldmath{$S$}}_{2n}$ 
and $\mbox{\boldmath{$S$}}_{2n+1}$ ($n=1,2,...$) are 
strengthened  (see eq.(\ref{5-0})) and 
dimers are formed between them, i.e.,
\begin{eqnarray}
\mbox{\boldmath{$S$}}_{1}-\mbox{\boldmath{$S$}}_{2}=
\mbox{\boldmath{$S$}}_{3}-\cdots
-\mbox{\boldmath{$S$}}_{N-2}=\mbox{\boldmath{$S$}}_{N-1}
-\mbox{\boldmath{$S$}}_{N}\nonumber 
\end{eqnarray} 
where $\mbox{\boldmath{$S$}}_{i}=\mbox{\boldmath{$S$}}_{i+1}$ 
represents the singlet bond. 
Then, in the case of the o.b.c., the spin at $j=1$ and 
that of $j=N$ {\it cannot} 
form a spin singlet due to the 
absence of $\mbox{\boldmath{$S$}}_{1}\cdot
\mbox{\boldmath{$S$}}_{N}$ term.
Namely, in the limit of $J=R$, the following four states
would be degenerated, 
\begin{eqnarray} 
&&|\downarrow\rangle_{1}\otimes({\rm{dimer\ chain}})
\otimes|\uparrow\rangle_{N},\label{5-1} \\
&&|\uparrow\rangle_{1}\otimes({\rm{dimer\ chain}})
\otimes|\downarrow\rangle_{N},\label{5-2} \\
&&|\downarrow\rangle_{1}\otimes({\rm{dimer\ chain}})
\otimes|\downarrow\rangle_{N},\label{5-3} \\
&&|\uparrow\rangle_{1}\otimes({\rm{dimer\ chain}})
\otimes|\uparrow\rangle_{N}.\label{5-4}
\end{eqnarray}
 The last two states ((\ref{5-3}),(\ref{5-4})) do not belong to the 
eigenspace of $S^{z}_{\rm{tot}}=0$ 
and we ignore them henceforth.\cite{comm4} When we introduce 
finite $J - R>0$, the $\frac{N}{2}$-th order perturbation 
 in terms of $\hat{S}^{+}_{1}\hat{S}^{-}_{2}$,
$\hat{S}^{+}_{3}\hat{S}^{-}_{4}$,..., 
and $\hat{S}^{+}_{N-1}\hat{S}^{-}_{N}$ (and their Hermite 
conjugate) lifts this degeneracy. Then the resulting gap 
should be scaled as $\exp(-N\alpha/l)$,  
where $l$ is expected to be a magnetic 
correlation length.  In fact, we can fit the size 
dependence of $\Delta_{1}$  
by this exponential function as in the inset of Fig. \ref{3}b, 
from which the correlation length at $\varphi=\frac{3\pi}{2}$ 
 is estimated around $3$ sites.  Because of this size dependence, 
we expect that $\Delta_{1}$ would vanish when we took the system size
sufficiently large compared with this correlation length.
\begin{figure}[t]
\begin{center}
\includegraphics[width=0.4\textwidth]{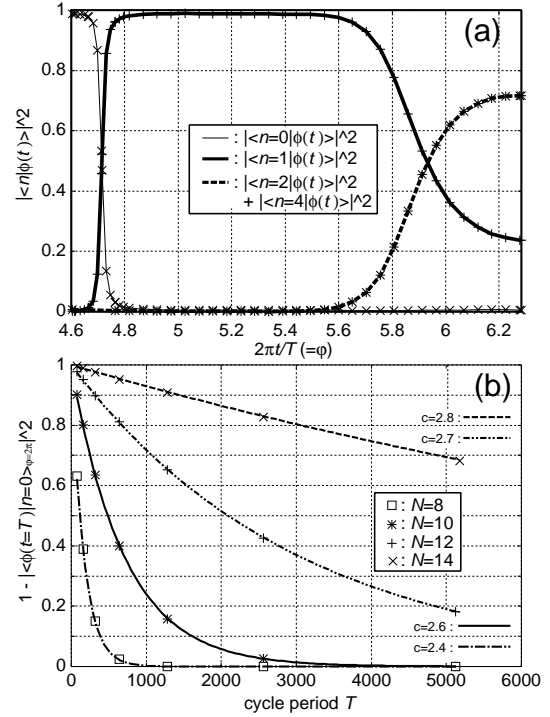}
\end{center}
\caption{(a): Projected weight of
 $|\phi(t)\rangle$ onto each instantaneous eigenstate 
$|\langle n|\phi(t)\rangle|^{2}$ for $n=0,1,2,4$ 
where the w.f. $|\phi(t)\rangle$ is always almost 
perpendicular to the excited state with $n=3$:
$|\langle n=3|\phi(t)\rangle|^{2}<10^{-7}$. 
$J=1.5$,$R=0.3$ and $N=14$. (b): 
Transition probability $P_{n=0\rightarrow n=1}$ 
as a function of cycle period $T$. Every data points 
are fitted by the Landau-Zener function 
$\exp\bigl(-\frac{\pi\Delta_{1}^{2}}{c\times2\pi/T}\bigr)$, 
where $c$ is estimated around $2.5$. $J=1.5$ and $R=0.3$.}
\label{4} 
\end{figure}

Because of this crossing character of the energy spectrum at 
$\varphi=\frac{3\pi}{2}$, the system which has resided on 
the ground state ($n=0$) transits, at $\varphi=\frac{3\pi}{2}$, 
into the first excited state 
with $n=1$ which is represented by the bold solid line 
in Fig. \ref{3}b. We can see this transition in Fig. \ref{4}a,  
where the projected weights of $|\phi(t)\rangle$  
onto each instantaneous eigenstate are given as a function 
of $\varphi=\frac{2\pi t}{T}$. 
Futhermore, its transition probability, $P_{n=0\rightarrow n=1}$, 
is well fitted by the Landau-Zener formula\cite{zener,miyashita} 
\begin{eqnarray}
P_{n=0\rightarrow n=1}=\exp(-\frac{\pi\Delta_{1}^2}
{c\times2\pi/T}), \nonumber 
\end{eqnarray}
as in Fig. \ref{4}b, where we plotted  
$P_{n=0\rightarrow n=1}$ for various cycle period $T$ 
and various system size. From this fitting, $c$ 
in the Landau-Zener function is estimated 
around $2.6 \pm 0.2$.\cite{comm5}

As shown in the Fig. \ref{4}a, after the first transition  
took place at $\varphi=\frac{3\pi}{2}$, the second transition 
 from the first excited state to the second excited state with $n=2$
 (represented by the bold broken line) happens at $\varphi\simeq 5.9$. 
However its transition probability $P_{n=1\rightarrow n=2}$ is 
 around $70\%$ and not so perfect compared with 
$P_{n=0\rightarrow n=1}$. This is mainly 
because the gap $\Delta_{2}$ around $\varphi\simeq 5.9$
between the first excited state and the second excited state 
is a substantial amount compared with $\Delta_{1}$ (see Fig. \ref{3}b).
However, as in the inset of Fig. \ref{3}b, its size dependence 
is also scaled by 
\begin{eqnarray}
\Delta_{2} \sim {\rm exp}(-N\alpha/l), \label{5-5} 
\end{eqnarray}
where the magnetic correlation length $l$ is estimated around 4.6 
sites. Therefore, when we take the system size large enough,  
we can naturally expect that the gap $\Delta_{2}$ would 
reduce exponentially, which enhances the transition probability 
$P_{n=1\rightarrow n=2}$ drastically. 
\begin{figure}[t]
\begin{center}
\includegraphics[width=0.4\textwidth]{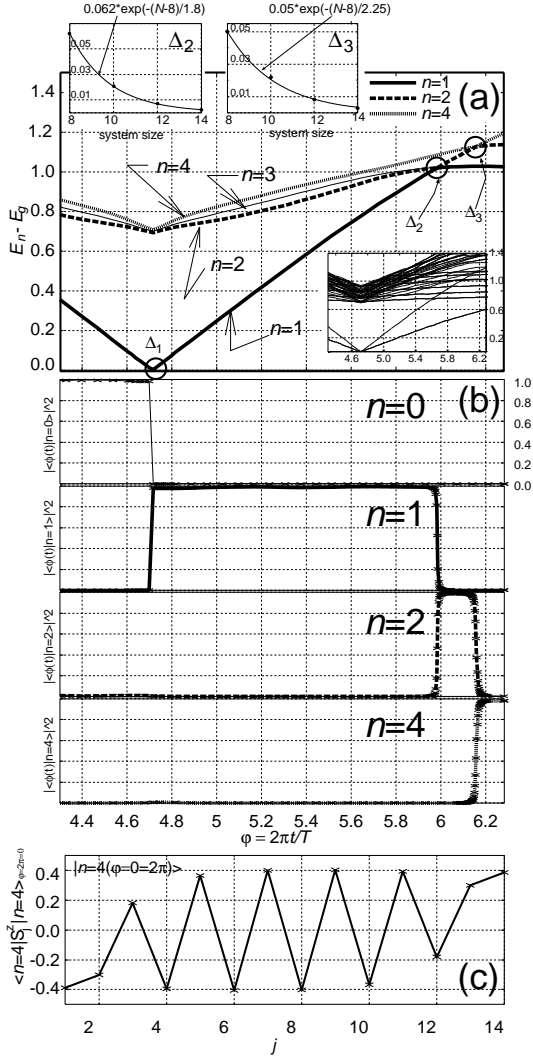}
\end{center}
\caption{ $J=0.6,R=0.3$ and $N=14$. 
(a): Energy spectrum for the eigenstates 
with $\hat{S}^{z}_{\rm{tot}}=0$ as a function 
of $\varphi$,(b): Projected weight of $|\phi(t)\rangle$ 
onto the lower five eigenstates, i.e.,
$|\langle \phi(t)|n\rangle|^2(n=0,1,2,4)$, (c): Expectation value of 
$\hat{S}^{z}_{j}$ taken over the fourth excited state with $n=4$  
$|n=4\rangle_{\varphi=2\pi}$}
\label{5} 
\end{figure}

In order to verify this expectation, we perform another numerical 
calculations. As it is very difficult to calculate numerically 
with larger system size ($N > 14$), we, instead, decrease the ratio  
$J/R$ in order to {\it{reduce the magnetic correlation
 length $l$}}, which is also expected to reduce these gaps 
according to eq.(\ref{5-5}). The result for $J=0.6,R=0.3$ 
and $N=14$ is summarized in Fig. \ref{5}. Figure\ref{5}a 
indicates the energy spectrum for the lower four excited states  
with $\hat{S}^{z}_{\rm{tot}}=0$ and Fig. \ref{5}b shows 
the projected weight, 
$|\langle n|\phi(t)\rangle|^{2}$ for $n=0,1,2,4$. 
Figure\ref{5}c represents the expectation value of $\hat{S}^{z}_{j}$ 
taken over the $|n=4\rangle_{\varphi=2\pi\equiv0}$ as a function of 
site index $j$. Then, as is expected, 
the gaps we observed in Fig. \ref{3} ($\Delta_{1},\Delta_{2}$) 
decrease drastically. In addition to them, there 
appears a new gap $\Delta_{3}$ between the excited state with  
$n=2$ and that with $n=4$ at $\varphi=6.15$. This gap $\Delta_{3}$ 
is also very tiny and scales as $\exp(-N/2.25)$. 

As a result of these crossing characters of energy spectrum at 
$\varphi=\frac{3\pi}{2},5.97$ and $6.15$, three Landau-Zener type 
tunnelings take place as shown in Fig. \ref{5}b.
 Namely, the wavefunction $|\phi(t)\rangle$ changes its weight 
from the ground state with $n=0$ to the first excited state with $n=1$ 
(represented by the bold solid line in Fig. \ref{5}a) at 
$\varphi=\frac{3\pi}{2}$ , from the first excited  state to the excited 
state with $n=2$ (bold broken line) at $\varphi=5.97$ and lastly  
from the excited state with $n=2$ to the fourth 
excited state with $n=4$ (bold dotted line) 
at $\varphi=6.15$. All these transition probabilities are 
almost $100\%$ in accordance with our previous 
expectations.\cite{comm6} 

After all, through these transitions,  
the wavefunction is raised from the ground state onto 
the fourth excited state ($n=4$) during this  
last quarter of the cycle. When we see the expectation value 
of $\hat{S}^{z}_{j}$ taken over this fourth excited state 
at $\varphi=0\equiv2\pi$ (see Fig. \ref{5}c), it is almost identical to 
the solid line of the lower panel in Fig. \ref{2}, where  
the total $\hat{S}^{z}$ on the left end
($\sum_{j=1}^{3}\langle \hat{S}^{z}_{j}\rangle$) decreases by 0.9 while 
that of the right end ($\sum_{j=12}^{14}\langle \hat{S}^{z}_{j}\rangle$) 
increases by 0.9 in comparison with that of the initial ground state 
wavefunction.

To summarize, the numerical observations found in Fig. \ref{2} can be 
ascribed to the difference between the energy spectrum 
with the p.b.c. and that with the o.b.c. In the latter case, 
a certain excited state decreases its energy level until  
$\varphi=\frac{3\pi}{2}$ and then {\it{picks up}} the system 
which has been staying on the ground state.
Then, through the Landau-Zener tunnelings, several excited states 
carry the system hand in hand perfectly upon a particular excited 
state at $\varphi=0$, whose $\hat{P}_{S^{z}}$  increases by 
almost $1$ in comparison with that of the initial state. 
These observations remain  unchanged if the sweeping 
velocity is always higher than  
$\frac{\pi\Delta^{2}_{i}}{c}$;
\begin{eqnarray}
\frac{1}{T} > \max_{i=1,2,3}
\Bigl(\frac{\pi\Delta^{2}_{i}}{c}\Bigr).
\end{eqnarray} 
This lower limit for the sweeping velocity is expected to vanish  
in the thermodynamic limit, since the $\Delta_{1,2,3}$ 
reduces {\it{exponentially}} to zero as the system size $N$
 becomes larger. 
Therefore the conclusion in eq. (\ref{4-1-0}) is consistent 
with the numerical results of this section.

Finally let us mention about the relevance of our story to real systems 
such as Cu-benzoate and ${\rm{Yb}}_{4}{\rm{As}}_{3}$. There 
the effective staggered field along the $z$-direction accompanies 
the uniform effective Zeeman fields along the $x$(or $y$)-direction, 
$H^{x(y)}_{u}$. 
Furthermore, since these systems break the bond-centered inversion 
symmetry, there is a Dzyaloshinskii-Moriya (DM) 
interaction.  Its DM vector must alternate bond  
by bond in the absence of the electric field. In addition to 
them, the uniform component of the DM vector would be also induced 
by the applied electric field in general. Even if these  
 terms are gradually introduced to our model calculations, 
 $\delta \theta_{+}$ remains invariant as far as 
the gap along the loop remains open, in other words, 
the system along the loop remains locked by the sliding potential 
$\sin(\hat{\theta}_{+}+\varphi(t))$.  This is a manifestation of the fact 
that the quantized $\delta \theta_{+}$ (in the thermodynamic limit) 
is a topologically protected quantity. However, detail studies 
on the effect of these terms 
might be remaining interesting problems. 

When the system travels around this critical ground 
state $M$ times, total $z$-component of spin around one end 
increases by $M$
, while that on the other end 
decreases by a same amount. 
Such an inhomogeneous magnetic structure 
can be detectable around  sample boundaries 
and/or magnetic domain boundaries by using 
some optical probes. However, we must 
mention that the excited  state with $P_{S^{z}}=m$ 
(In the case of $m=1$, this corresponds to $|n=4\rangle_{\varphi=0}$ 
in Fig. \ref{5}) might fall into the ground state 
$|n=0\rangle_{\varphi=0}$ by way 
of the $x$-component of the uniform magnetic field $H^{x}_{\rm{u}}$ 
at around $t=mT$, when we sweep the system slower 
than $2\mu_{B}H^{x}_{\rm u}$. In order to avoid this relaxation process, 
the sweeping velocity should be taken faster than
$2\mu_{B}H^{x}_{\rm{u}}$. 
\begin{eqnarray}
\frac{1}{T} > 2\mu_{B}H^{x}_{\rm u},
\end{eqnarray}
However, in order to keep the system on a particular 
minimum of the locking potential 
$\cos(\theta_{+}+\frac{2\pi t}{T})$, the velocity 
must be also slower than the spin gap observed 
along the loop,
\begin{eqnarray}
\frac{1}{T} < J^{1/3}R^{2/3}.
\end{eqnarray} 
Therefore we need to have a finite window between the 
lower limit of the sweeping velocity and the upper limit.
In the case of Cu-benzoate, the spin gap of $0.1\rm{meV}$ 
is induced by the applied magnetic field 
of $1\rm{T}$ $(2\mu_{B}\times 1{\rm T}\sim 0.1\rm{meV})$. 
Thereby, the lower limit of this sweeping velocity 
is unhappily comparable to the higher limit in this system. 
However, spin gaps around $0.1\rm{meV}$ could be also 
induced by the magnetic field {\it smaller} than $1\rm{T}$ 
in those quantum spin chains with {\it relatively larger} $J$, 
where substantial amount of the window between the upper 
limit and the lower limit would exist. 
Therefore, it is not so hard to 
realize our story experimentally on the quasi-1D quantum
 spin systems with relatively large intrachain
 interaction and with two crystallographically 
inequivalent sites in its unit cell.    

We also want to mention about the magnitude of the electric field 
required in order to induce the dimerization gap enough to be 
observable and also enough to be comparable to the spin gap induced 
by the magnetic field, which is around $0.1{\rm{meV}}$.
Since it goes beyond the scope of this paper to quantify 
microscopically the change of the exchange interaction induced 
by external electric fields, we will pick up 
some reference data from 
other materials. Let's see magneto-electric materials, 
where the applied electric field often changes its 
effective exchange interactions $J$ due to its peculiar 
crystal structure and its change $\Delta$ could be 
quantitatively estimated via the resulting magnetization.
\cite{hornreich}  
In the case of famous magneto-electric material 
${\rm{Cr}}_{2}{\rm{O}}_{3}$, the intra-sublattice
exchange interaction for the + sublattice and that for the $-$ 
sublattice should be equal to each other because of the inversion 
symmetry ($J_{+}=J_{-}=J$). Then an electric field applied along 
its principle axis breaks this symmetry and induce a finite difference 
$\Delta=J_{+}-J_{-}$, where $\Delta/J$  was estimated around 
$10^{-5}$ under $E\sim 1{\rm{kV/cm}}$.\cite{hornreich} Namely,
\begin{eqnarray}
\frac{\Delta}{J} = c E. 
\end{eqnarray} 
with its linear coefficient $10^{-5}[{\rm cm/kV}]$. 
When we apply this coefficient to our Heisenberg model, 
eq.(\ref{-1-0-1}) with $J\sim 0.1\rm{eV}$, 
we need an electric field 
of order of $1{\rm{kV/cm}}$ in order to induce the 
dimerizing gap $J^{1/3}\Delta^{2/3}$ of order of $0.1\rm{meV}$, 
which is still one order of magnitude smaller than the typical 
value of the Zener's 
breakdown field ($\ge10^{4}\rm{V/cm}$ at $T\le 100{\rm K}$) 
of $1$D Mott insulators.\cite{taguchi1}
\section{Conclusion and discussion}   
In this paper, we propose the quantum spin pumping 
in $S=1/2$ quantum spin chains with two crystallographically 
inequivalent sublattices in its unit cell ($T, I_{\rm{bond}}$ : broken 
and $I_{\rm{site}}$ : unbroken) and with relatively large intrachain 
exchange interactions.  
Due to its peculiar crystal structure, an 
applied electric field ($E$) and magnetic field ($H$) endow 
its critical ground state with the finite spin gap 
via the dimerizing field ($\Delta\sim E$) and the
 staggered magnetic field ($h_{\rm{st}}\sim H$), respectively,
 where the phase operator $\hat{\theta}_{+}$ in sine-Gordon theory
 is locked on a particular valley of the relevant potential      
$\sqrt{h_{\rm{st}}^{2}+\Delta^{2}}\sin(\hat{\theta}_{+}+\varphi)$ : 
$\langle \hat{\theta}_{+} \rangle = \frac{\pi}{2}-\varphi+2\pi n$. 
When the system is deformed slowly along the loop which 
encloses the critical ground state in the $E$-$H$ plane,  
the  locking potential  
$\sqrt{h_{\rm{st}}^{2}+\Delta^{2}}\sin(\hat{\theta}_{+}+\varphi)$ 
in the sine-Gordon model slides gradually (see Fig. \ref{1}). 
After one cycle along this loop,  
the expectation value of the phase 
operator $\hat{\theta}_{+}$ for the ground state acquires $2\pi$ holonomy. 
This means that a spatial polarization of $z$-component 
of spins $P_{S^{z}}$  increases by $1$ after this cycle \cite{sharma}. 
In other words, the $z$-component of the spin current quantized 
to be $1$ flows through the bulk system during this cycle. 
 
These arguments are supported by the numerical analyses, 
where we performed the exact diagonalization of the 
finite size system and developed the ground state 
wavefunction along this loop in a thermally insulating way.

\begin{acknowledgments}
The author acknowledges A. V. Balatskii, 
Masaaki  Nakamura, M. Tsuchiizu, 
K. Uchinokura, T. Hikihara and 
A. Furusaki, S. Miyashita, 
S. Murakami and N. Nagaosa 
for their fruitful discussions and for their critical 
readings of this manuscript. 
The author is very grateful to K. Uchinokura 
for his critical discussions which are essential 
for this work. 
\end{acknowledgments}

\appendix
\section{Equivalence between the phase operator and the spatial
 polarization operator}
In this appendix, we will show that the continuum limit 
of $\hat{P}_{S^{z}}$ corresponds to the phase operator 
$-\frac{1}{N\alpha}\int\frac{\hat{\theta}_{+}(y)}{2\pi}{\rm d}y$, where  
 related work was done by Nakamura and Voit.\cite{nakamura} 
Historically the polarization operator is known to be an ill-defined 
operator on the Hilbert space with the periodic 
boundary condition (p.b.c.) and is consequently formidable to treat 
in solid state physics. However, 
recently King-Smith and Vanderbilt\cite{vanderbilt,ortiz}
 and Resta\cite{resta,resta0,resta1,aligia}
 showed that the {\it{derivative}} of 
the polarization with respect to some external parameter $\lambda$ 
can be expressed by the {\it{current operator}} which is in turn 
well-defined in the Hilbert space with the p.b.c.. 
Based on this observation, they quantitatively estimated   
electronic contributions to macroscopic polarizations  
in dielectrics and semiconductors such as GaAs, ${\rm{KNbO}}_{3}$ 
and III-V nitride. Following their strategies, we will 
show the equivalence between the 
derivative of 
$-\frac{1}{N\alpha}\int {\rm d}y \langle 
\frac{\hat{\theta}_{+}(y)}{2\pi} \rangle$ 
with respect to $\lambda$ and that of $\langle P_{S^{z}} \rangle$.

According to the standard perturbation theory\cite{resta0}, 
the derivative of the polarization with respect to some 
mechanical parameter $\lambda$ is given by 
\begin{eqnarray}
&&\frac{\partial \langle \hat{P}_{S^{z}}\rangle}{\partial \lambda}
=-\sum_{N\ne g}\nonumber \\
&&\Bigl[\frac{\langle g(\lambda)|[{\hat{P}_{S^{z}}}
,\hat{H}]| N(\lambda) \rangle \langle N(\lambda)|
\frac{\partial \hat{H}(\lambda)}{\partial \lambda}
|g(\lambda)\rangle}{(E_{g}(\lambda)-E_{N}(\lambda))^2}+{\rm{c.c.}}\Bigr], 
\label{2-1}
\end{eqnarray}
where we assume the system has a finite spin gap from the ground state.
The time derivative of $P_{S^{z}}$ in the numerator reduces to the 
summation of the current operator over all bonds; 
\begin{eqnarray} 
&&-{\rm i}\Bigl[{\hat{P}_{S^{z}}},H\Bigr] = 
-{\rm i}\Bigl[{\hat{P}_{S^{z}}},H_{\rm XY} + H_{\rm dim} \Bigr]
\nonumber \\ 
&&={\rm i}\frac{1}{N}\sum_{j=1}^{N}\sum_{k=1}^{N-1} 
\Bigl[j\hat{S}^{z}_{j}, -\frac{1}{2}\bigl[J+(-1)^{k}\Delta\bigr]
(\hat{S}^{+}_{k}\hat{S}^{-}_{k+1}+{\rm{c.c.}})
\Bigr]\nonumber \\
&&=\frac{\rm i}{2N}\sum_{j=1}^{N-1}\bigl[J+(-1)^{j}\Delta\bigr]
(\hat{S}^{-}_{j}\hat{S}^{+}_{j+1}-\hat{S}^{+}_{j}\hat{S}^{-}_{j+1}).
\label{3-0}
\end{eqnarray}
Thereby the contribution of every single bond is 
always ${\cal{O}}(1/N)$, which is not the case with 
$\hat{P}_{S^{z}} = \frac{1}{N}\sum_{j=1}^{N} j \hat{S}^{z}_{j}$. 
As a result, we are allowed to estimate   
{\it with the periodic boundary condition} the r.h.s. of eq. (\ref{3-0}) 
in the thermodynamic limit, which we should have figured out 
with the open boundary. 
This is because their difference is attributed
to the spin currents on the several bonds\cite{comm2d} around the
boundaries, which is at most ${\cal{O}}(1/N)$ in our 
spin gapped system. Then we will consider 
eq. (\ref{3-0}) with the p.b.c. and 
rewrite it in terms of the bosonization language. 
That is to say, we express the current operator in  eq. (\ref{3-0}) 
in terms of the phase operator $\hat{\theta}_{+}(x)$ and 
its conjugate field $\hat{\Pi}(x)$, 
\begin{eqnarray}
&&\hat{S}^{+}_{j+1}\hat{S}^{-}_{j}-\hat{S}^{+}_{j}\hat{S}^{-}_{j+1}=
f^{\dagger}_{j+1}f_{j}-f^{\dagger}_{j}f_{j+1}\nonumber \\
&&\approx\  \Bigl[R^{\dagger}(x_{j+1})e^{-{\rm i}k_{F}x_{j+1}} 
+ L^{\dagger}(x_{j+1})e^{{\rm i}k_{F}x_{j+1}}\Bigr] \nonumber \\
&&\hspace{0.5cm} \times \ \Bigl[R(x_{j})e^{{\rm i}k_{F}x_{j}} 
+ L(x_{j})e^{-{\rm i}k_{F}x_{j}}\Bigr] - {\rm H.c.} \nonumber \\ 
&&\approx\ -2{\rm i}\cdot\Bigl[R^{\dagger}(x_{j})R(x_{j}) 
- L^{\dagger}(x_{j})L(x_{j})\Bigr]  \nonumber \\
&&+\  {\rm i}\alpha\cdot(-1)^{j}\Bigl[R^{\dagger}(x_{j})
{\partial_x}L(x_{j})-L^{\dagger}(x_{j})
{\partial_x}R(x_{j}) + {\rm{H.c.}}\Bigr]\nonumber \\
&&=\ 4{\rm i}\hat{\Pi}(x_{j}) \nonumber \\
&&\ \  +\ {\rm i}(-1)^{j}\Bigl[2\hat{\Pi}(x_{j})\cos\hat{\theta}_{+}(x_{j})
+ {\rm H.c.} 
\Bigr]. \label{3-1}
\end{eqnarray}
Accordingly, the time derivative of $P_{S^{z}}$ in the 
continuum limit can be expressed in the following form,  
\begin{eqnarray}
&&-{\rm i}\Bigl[{\hat{P}_{S^{z}}},H\Bigr] \nonumber \\
&&\approx -\frac{1}{N\alpha}
\int {\rm d}x \biggl\{2J\hat{\Pi}
 + \Delta\Bigl[\hat{\Pi}\cos\hat{\theta}_{+} 
+ {\rm H.c.} \Bigr]\biggr\}. \label{3-2}  
\end{eqnarray}  

From now on, we will show that this expression is 
identical to the time derivative of the phase operator: 
\begin{eqnarray}
\mbox{(r.h.s. of eq.(\ref{3-2}))}
={\rm i}[\frac{1}{N\alpha}\int {\rm d}x
\frac{\hat{\theta}_{+}(x)}{2\pi},\hat{H}]. 
\end{eqnarray}
The first term in the r.h.s. of  eq.(\ref{3-2}) comes from 
the commutator between $\hat{\theta}_{+}$ and $H_{\rm XY}$ 
given in eq.(\ref{2-0-3}), 
\begin{eqnarray}
{\rm i}\Bigl[\frac{\hat{\theta}_{+}(x)}{2\pi},H_{\rm{XY}}\Bigr]
=-2J\hat{\Pi}(x). \label{3-2-0}
\end{eqnarray}
On the other hand, the commutator between the phase operator 
and $H_{\rm dim}$ given in eq.(\ref{2-0-3-2}) vanishes and does 
not produce the last two terms of eq. (\ref{3-2}). 
This is because eq.(\ref{2-0-3-2}) is not an accurate expression
for $H_{\rm dim}$ in the continuum 
limit and needs some additional terms, whose commutators with 
the phase operator correctly yield the last two terms 
of eq. (\ref{3-2}).
In order to prove that this is indeed the case, we will 
carefully reexamine the bosonization of  
$\hat{S}^{+}_{j}\hat{S}^{-}_{j+1}+\hat{S}^{+}_{j+1}\hat{S}^{-}_{j}$ 
\cite{comm2e} :  
\begin{eqnarray} 
&&\hat{S}^{+}_{j}\hat{S}^{-}_{j+1} + {\rm H.c.} \nonumber \\
&&=\  \Bigl[R^{\dagger}(x_{j})e^{-{\rm i}k_{F}x_{j}} 
+ L^{\dagger}(x_{j})e^{{\rm i}k_{F}x_{j}}\Bigr] \nonumber \\
&&\hspace{0.5cm}\times\ \Bigl[R(x_{j+1})e^{{\rm i}k_{F}x_{j+1}} 
+ L(x_{j+1})e^{-{\rm i}k_{F}x_{j+1}}\Bigr] + {\rm H.c.} \nonumber \\ 
&&\approx\ ({\rm{r.h.s}}\ \ {\rm{of}} \ \ {\rm{eq. (\ref{1-1-2})}}) 
 -{\rm i}(-1)^j\frac{\alpha^{2}}{2}\cdot\Bigl[R^{\dagger}(x_{j})
{\partial_x}^{2}L(x_{j}) \nonumber \\
&&\ \ \  - L^{\dagger}(x_{j}){\partial_x}^{2}R(x_{j}) - {\rm
 H.c.}\Bigr] \nonumber \\
&&= \mbox{ (r.h.s. of  eq. (\ref{2-0-2}))} \nonumber \\
&&\hspace{0.5cm} +\ 
(-1)^{j}\alpha\biggl\{-\frac{1}{4\pi}(\partial_{x}\hat{\theta}_{+}(x_j))^{2}
\cos\hat{\theta}_{+}(x_{j}) \nonumber \\
&&\hspace{0.5cm} +\ 
\pi\Bigl[\hat{\Pi}^{2}(x_{j})\cdot\cos\hat{\theta}_{+}(x_j) \nonumber \\
&&\hspace{0.5cm} +\ \hat{\Pi}(x_j)\cdot\cos\hat{\theta}_{+}(x_j)\cdot\hat{\Pi}(x_j) 
+  {\rm{H.c.}}\Bigr]\biggr\}.\label{3-2-1}
\end{eqnarray}
Then, by taking its staggered components, 
we obtain the following expression for $H_{\rm dim}$, instead of 
eq.(\ref{2-0-3-2}), 
\begin{eqnarray}
&&H_{\rm{dim}}= \nonumber \\
&&\Delta \int {\rm d}x \Biggl\{-\frac{1}{\pi\alpha^{2}}
 \cos\hat{\theta}_{+} 
- \frac{1}{8\pi}\left(\partial_{x}\hat{\theta}_{+}\right)^{2}
\cos\hat{\theta}_{+} \nonumber \\
&&  
+\ \frac{\pi}{2}\biggl[\hat{\Pi}^{2}\cdot\cos\hat{\theta}_{+}
+ \hat{\Pi}\cdot\cos\hat{\theta}_{+}\cdot\hat{\Pi}
+{\rm{H.c.}}\biggr]\Biggr\}. \label{3-2-2}  
\end{eqnarray}
Here we want to mention that, irrespective of 
this modification, $\cos\hat{\theta}_{+}$ in $H_{\rm dim}$ 
always locks the phase operator in combination with $\sin\hat{\theta}_{+}$ 
in $h_{\rm st}$ as far as $|\gamma| \leq 1$ (see the text). 
This is because the RG eigenvalues of the additional terms such as 
$(\partial_{x}\hat{\theta}_{+})^2\cos\hat{\theta}_{+}$,  
$(\hat{\Pi})^{2}\cos\hat{\theta}_{+}$ and etc. are all negative  
and thereby irrelevant in the sense of the renormalization group study. 
However these additional terms in $H_{\rm dim}$ cannot be 
discarded, since the commutator between these terms 
and the phase operator produces the last two term of eq.(\ref{3-2}): 
\begin{eqnarray}
&&{\rm i}\bigl[\frac{\hat{\theta}_{+}(x)}{2\pi}, H_{\rm dim} \bigr]\nonumber \\
&&=\ -\Delta \biggl[\hat{\Pi}(x)\cos\hat{\theta}_{+}(x) 
+ \cos\hat{\theta}_{+}(x)\hat{\Pi}(x)\biggr].\label{3-2-3}
\end{eqnarray}

After all, by comparing eqs.(\ref{3-2-0}) and (\ref{3-2-3}) with 
eq.(\ref{3-2}), we can safely replace $-i[P_{S^z},H]$ by 
$\frac{i}{N\alpha}\int {\rm d}x[\frac{\hat{\theta}_{+}}{2\pi},H]$ in the 
continuum limit and  rewrite eq. (\ref{2-1}) into  
a following simple form by using the bosonization language:
\begin{eqnarray}
\frac{\partial \langle \hat{P}_{S^{z}}\rangle}{\partial \lambda}
&\approx&\frac{1}{N\alpha}\sum_{N\ne g}\int {\rm d}x \Biggl[\frac{\langle g|
[\frac{\hat{\theta}_{+}(x)}{2\pi},\hat{H}]
| N \rangle \langle N|\frac{\partial \hat{H}}
{\partial \lambda}|g\rangle}{(E_{g}-E_{N})^2} + {\rm{c.c.}}\Biggr]\nonumber \\
&=&-\frac{1}{N\alpha}
\sum_{N}\int {\rm d}x \Bigl[\langle g|\frac{\hat{\theta}_{+}(x)}{2\pi}
| N \rangle \langle N|\frac{\partial}{\partial \lambda}|g\rangle
 + {\rm{c.c.}}\Bigr]\nonumber \\
&=&-\frac{1}{N\alpha}\int {\rm d}x
\frac{\partial}{\partial \lambda}\langle\frac{\hat{\theta}_{+}(x)}{2\pi}
\rangle.\label{4-1}
\end{eqnarray} 
Here we used $\langle N|\frac{\partial \hat{H}}{\partial \lambda}|g\rangle
=(E_{g}-E_{N})\langle N|\frac{\partial}{\partial \lambda}|g\rangle$.


\begin{thebibliography}{9}
\bibitem{comm0} Hamiltonian has also a staggered Dzyaloshinskii-Moriya vector, 
whose effect can be renormalized into the staggered $g$-tensor 
$[g_{\rm{a}}]$. When the system is invariant under the 
$\pi$-rotational symmetry which exchanges nearest neighbor sites, bare $[g_{a}]$ 
becomes a non-generic matrix in general.
\bibitem{comm2} If the system is invariant under the $\pi$-rotational 
symmetry which exchanges the nearest neighboring {\it{bonds}}
 , the direction of  $\mbox{\boldmath{$E$}}$ must be deviated from this 
$\pi$-rotational axis.
\bibitem{comm1-1} Here we also rotate by $\pi$ around the $z$-direction 
in the spin space on odd sites:
\begin{eqnarray}
(\hat{S}^{x}_{2j+1},\hat{S}^{y}_{2j+1},\hat{S}^{z}_{2j+1})
\rightarrow (-\hat{S}^{x}_{2j+1},-\hat{S}^{y}_{2j+1},\hat{S}^{z}_{2j+1})
\end{eqnarray}
We take this sign convention only in $\S$ 2.
\bibitem{comm4}The energy level which reaches around 0.9 
at $\varphi=2\pi$ in Fig.\ref{3}a corresponds to 
doubly degenerate eigenstates which belong to the 
eigenspaces of $\hat{S}^{z}_{\rm{tot}}=\pm1$ respectively. 
These doubly degenerate states at $\varphi=\frac{3\pi}{2}$ 
corresponds to the two state given in eqs. (\ref{5-3}) and 
(\ref{5-4}). On the contrary, the eigenstate labeled as $n=1$ 
(represented by the bold solid line in Fig. \ref{3}b) 
and the ground state belong to the eigenspace 
of $\hat{S}^{z}_{\rm{tot}}=0$. Since our time-dependent 
Hamiltonian conserves the total spin density $S^{z}_{\rm{tot}}$,  
$|\phi(t)\rangle$ does not have projected weights onto 
those eigenstates with $S^{z}_{\rm{tot}}\ne 0$.
\bibitem{comm6} Here the wavefunction does not have the 
projected weight on the third excited state with $n=3$ 
(thin line in Fig. \ref{5}a) during the cycle.
\bibitem{comm7}The uniform component of the Zeeman field does 
not cause the spin gap to the critical ground state of the 
isotropic Heisenberg model as far as it is small. 
In contrast the staggered component is always relevant even if 
it is infinitesimally small. However when the uniform 
component of the magnetic field $H_{u}$ is very strong, 
the system will be governed by the fixed point where the 
ground state is ferromagnetically polarized with a finite spin gap. 
This fixed point is different from the fixed point at which 
the system is locked by the staggered component of the 
magnetic field, i.e.,  $h_{\rm{st}}\sin\hat{\theta}_{+}$. 
As far as the system under the applied magnetic field 
is governed  by the latter fixed point, 
$\delta \theta_{+}$ is quantized to be $1$.
\bibitem{comm2d} In spin-gapped systems, the effects of 
the boundary range only within its magnetic correlation 
length from both ends.
\bibitem{comm2e} In the final line of eq.(\ref{3-2-1}), 
we drop those terms which can be summarized into a form of the 
total derivative.
\end{thebibliography}
\end{document}